\documentclass[a4paper,twoside,11pt,dvipsnames]{article}
\usepackage[utf8]{inputenc}

\usepackage[english]{babel}
\usepackage{subeqnarray}
\usepackage{url}

\usepackage{graphicx,psfrag}
\usepackage{fixltx2e,amsmath}

\usepackage{amssymb}
\usepackage[final]{pdfpages}
\usepackage{bbold}
\usepackage{dsfont}
\usepackage{pgfplots}
\usepackage{subfig}
\usepackage{algorithm}
\usepackage{algorithmic}
\usepackage[font=small,labelfont=bf]{caption}

\usepackage[font={footnotesize}]{caption}

\usepackage[utf8]{inputenc}
\usepackage[T1]{fontenc}
\usepackage{amsmath,amssymb}
\usepackage{graphicx}
\usepackage[sort&compress,numbers]{natbib}
\usepackage{enumerate}
\usepackage{wasysym}
\usepackage[textsize=footnotesize]{todonotes}
\usepackage{url}

\usepackage{pgfplots}

\definecolor{darkblue}{rgb}{0,0,0.6}
\definecolor{darkred}{rgb}{0.6,0,0}

\usepackage{hyperref}

\usepackage{authblk}

\usepackage[bitstream-charter]{mathdesign}

\usepackage{geometry}
\usetikzlibrary{pgfplots.groupplots}

\newcommand{\red}[1]{{#1}}

\graphicspath{{figures/}} 

\title{Equity Factors: To Short Or Not To Short, That Is The Question}

\author{Florent Benaych-Georges, Jean-Philippe Bouchaud, Stefano Ciliberti}
\affil{Capital Fund Management} 
\begin{document}

\date{\today}

\maketitle

\begin{abstract}
What is the best market-neutral implementation of classical Equity Factors? Should one use the specific predictability of the short-leg to build a zero beta Long-Short portfolio, in spite of the specific costs associated to shorting, or is it preferable to ban the shorts and hedge the long-leg with -- say -- an index future? We revisit this question by focusing on the relative predictability of the two legs, the issue of diversification, and various sources of costs. Our conclusion is that, using the same Factors, a Long-Short implementation leads to superior risk-adjusted returns than its Hedged Long-Only counterpart, at least when Assets Under Management are not too large. 
\end{abstract}

\tableofcontents


\section{Introduction}\label{introduction}

Equity Factor investing has become increasingly popular over the past
decade. Practitioners and academics realized in the 70's that the single factor, CAPM
model~\cite{Sharpe1964} has its own limitations and had to be
generalized in order to account for more than one risk
driver~\cite{Ross1976}. Several questions remain open regarding the nature of these extra factors:
are they pure risk premia~\cite{Zhang2006,Amihud2002,Chan1991, Vassalou2004},
genuine market anomalies~\cite{Sloan1996,DeBondt1987,Bouchaud2016,Baker2011}, or
unavoidable consequences of institutional constrained
investors~\cite{Dasgupta2011}? How many such factors must be considered: 3,
as in the original Fama-French model \cite{FamaFrench93}, or several hundreds as 
advocated in the so-called factor-zoo literature~\cite{HarveyLiu2013, HarveyLiu2019}?.  

There is also a wide variety of ways these factors can be converted into predictive signals and 
realistic portfolios. The primary criterion concerns the market exposure of the portfolio: are we 
looking for a market-neutral implementation of these factors, with the long and short legs of the portfolio offsetting its overall beta exposure, or are we concerned with so-called smart-beta strategies, where the portfolio has a positive beta exposure to the equity market, but is tilted in the direction of said factors? 

Both approaches make sense, of course, and correspond to different asset management mandates
and different investor profiles. Still, for any given portfolio we can always identify 
and isolate the portfolio`s exposure to the equity index and analyse the performance of its market neutral, active component. As an example, in the smart-beta style of implementation, this
market neutral component can be represented by a portfolio with long-only
equity positions, hedged by a short equity index futures. This is in fact a realistic, cost-aware set-up, which allows one to build a market-neutral factor strategy. 

An alternative is the classical Long-Short equity portfolio, with an explicit short position on some stocks. The question that we are going to address in this paper is whether any of these two
implementations yields significantly better results for the active, market-neutral risk component of the portfolio, in particular when realistic transaction costs are accounted for. Or, stated differently: {\it Are explicit short positions beneficial or detrimental to equity factor strategies?} 

Not surprisingly, we will see that the predictability power of the short signals 
plays a crucial role in determining which of the two implementations should be preferred. 
We will also show that a proper answer to this question depends quite heavily on some details 
related to the factor under scrutiny, on the portfolio construction algorithm, and on the very definition of the equity market. We will pay special attention to implementation costs, including both trading costs (impact) and financing costs, the latter substantially affecting the profitability of the short leg.  

This problem has been addressed in the literature many times in the past. 
Some authors focus on the difficulties related to actually taking short
positions~\cite{Bambaci}, while others find that under some well
defined market conditions the short leg of equity strategies is more
profitable~\cite{Stambaugh2011}. There seems to be a general agreement that
Long-Short strategies are more profitable than constrained long-only
ones~\cite{Ilmanen2012}, despite some claims that the extra-benefit is
marginal and that it would disappear when costs are properly accounted for~\cite{Robeco1}.

More recently, the authors of~\cite{Robeco2} have posted a study which suggests that 
an optimal implementation of market-neutral equity factors should not contain explicit 
short positions at all. Not only long signals seem to be of better quality than short signals, 
but also long positions provide a more diversified exposure to different factors than the shorts.

The contribution of the present study to the debate consists in proposing a realistic,
cost-aware framework allowing to compare a Long-Short equity portfolio
to a beta-Hedged Long-Only one. In our view, there are often two
missing ingredients in the previous research works that should
carefully be taken into account. One is the risk management and
the portfolio construction, that is a proper control of the portfolio
exposure to the desired descriptor under risk-related constraints --
which could be constant volatility and/or maximum leverage for
instance. This step is particularly crucial when considering long-only
portfolios, as we will show in Section \ref{parsconstruens}. The other
key point is costs. Allowing for short positions implies leverage and
borrowing costs that we will consider explicitly when building the
portfolio. 

The outline of the paper is as follows. In Section \ref{toymodel} we
introduce a simple toy model which allows one to rationalize when the use of 
shorts can bring value to a portfolio. When confronted to real predictability data, 
our criterion suggests that shorts are indeed accretive. In Section \ref{parsdestruens} we 
revisit the recent paper of the Robeco group~\cite{Robeco2}. While we reproduce the results of that study, we find that their conclusion that one should ``drop the shorts'' has to be tempered. As we show in 
Section \ref{parsconstruens}, a cost-aware portfolio construction using the full range of factor predictors leads to a clear over-performance of a Long-Short implementation compared to a beta-hedged long only portfolio.

\section{Building Intuition: Insights From A Toy Model}\label{toymodel}

\red{Consider an investment universe with a factor $F$, a market $M$ and  two assets, Asset 1, positively exposed to $F$ and Asset 2, negatively exposed to $F$ (both being positively exposed to the market).
The question is whether, to bet on $F$ without being exposed  to the market, we should rather bet on Asset 1 and hedge with a negative market exposure (\emph{hedged long-only}) or construct a market neutral portfolio with a bet on Asset 1 and a bet against Asset 2 (\emph{long-short}).}

  \red{We naturally model the respective returns $R_1$ and $R_2$ of Assets 1 and 2 as follows \begin{eqnarray}
  R_1 & = & \beta_1 M+ \alpha_1 F+\varepsilon_1 \ \ \ \text{ with $\alpha_1,\,\beta_1>0$,}\\
  R_2 & = & \beta_2 M - \alpha_2 F+\varepsilon_2 \ \ \  \text{ with $\alpha_2,\,\beta_2>0$,}
\end{eqnarray} where $\varepsilon_1$ and $\varepsilon_2$ 
 are 
idiosyncratic residuals.}  Up to a rescaling of both
assets, one can suppose without loss of generality that $\beta_1=\beta_2=1$. Then, up to a further
rescaling of $F$, one can set $\alpha_1=1,$ so that the above equations
rewrite
\begin{eqnarray}
  R_1 & =  & M+  F+\varepsilon_1\\
  R_2 & =  & M - \alpha_2 F+\varepsilon_2 \ . 
\end{eqnarray}
We shall also assume that:
\begin{itemize}
\item $\mathbb{E}( F ) >0$ (factor $F$ has  positive performance on average),
\item $F,\varepsilon_1,\varepsilon_2$ are independent and  $\mathbb{E}( \varepsilon_i)=0$ for $i=1,2$ (residual returns have no alpha).
\end{itemize}
We want to build a portfolio betting on $F$ (hence on Asset 1) in a
\emph{market-neutral way}, i.e. with no global exposure to $M$.  Up to a global rescaling of the positions, the two possibilities at hand are 
a Long-Short (LS) portfolio and a long position that we beta-hedge
(LH), that is:
\begin{eqnarray*}
\pi_{\text{LS}}  & \equiv & \{1,-1,0\}  \\
\pi_{\text{LH}}  & = & \{1,0,-1\} 
\end{eqnarray*}
where $\pi=$(respective weights of
  Asset  1, Asset 2 and Market). The following result compares the Sharpe Ratios of these portfolios
and determines, depending on the parameters $\alpha_2$,
$\gamma \equiv \text{Var}(\varepsilon_1)/\text{Var}(F)$ and $\kappa\equiv\text{Var}(\varepsilon_2)/\text{Var}(\varepsilon_1)$,
which one should be preferred.

\begin{figure} 
  \centering
  \includegraphics[scale=.38]{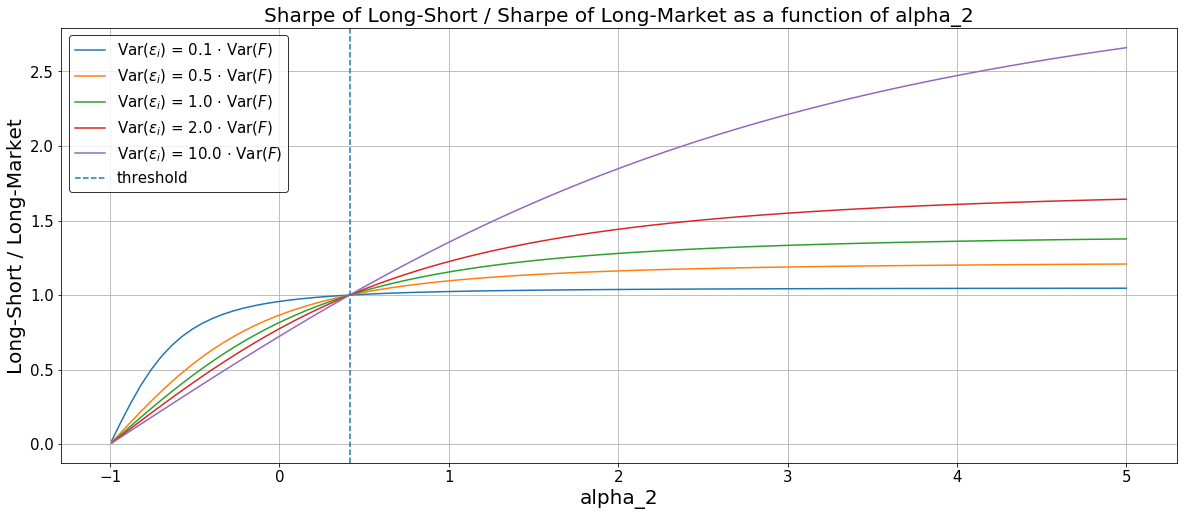}
  \caption{Illustration of \eqref{eq_ratio_sr} 
    when $\kappa=\text{Var}(\varepsilon_2)/\text{Var}(\varepsilon_1)=1$, for various values of
    $\gamma=\text{Var}(\varepsilon_1)/\text{Var}(F)$. When the variance of $F$ dominates
    that of the residues $\varepsilon_i$, (blue curve), the ratio of Sharpe
    ratios stays close to $1$ even for large values of $\alpha_2$,
    because in this case, both portfolios are approximately equivalent
    up to a global rescaling.}
  \label{fig_sr_ratio}
\end{figure} 

We find (see Appendix.~\ref{app1} for details) the following results:
\begin{equation}
  \label{eq_ratio_sr}
  \frac{\text{SR}(\pi_{\text{LS}})}{\text{SR}(\pi_{\text{LH}})} = \sqrt{\frac{1+\gamma}{1+\frac{\gamma(1+\kappa)}{(1+\alpha_2)^2}}}
\end{equation}
which implies that
\begin{eqnarray}
  \text{SR}(\pi_{\text{LS}})>\text{SR}(\pi_{\text{LH}}) &\iff& \alpha_2>\sqrt{1+\kappa}-1.
\end{eqnarray}
This simple result elicits a transition between two
regimes. If the \emph{hedging asset} (Asset 2 here) under-performs the
market enough, then from a (no-costs) Sharpe Ratio perspective the
long--short market-neutral portfolio on Assets 1 and 2 outperforms the
beta-hedged long portfolio on Asset 1. This is in line with the
intuition that, if we have enough predictability on the short part of
our portfolio, then shorts should be included. 

\begin{figure} 
  \centering
\includegraphics[width=.8\textwidth]{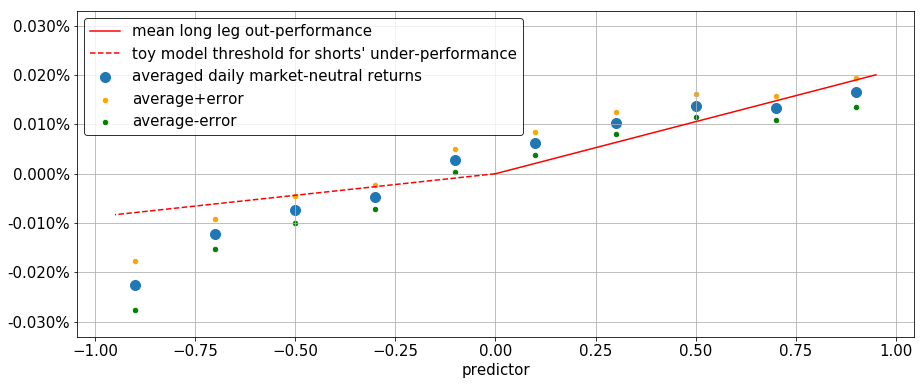}
  \caption{The predictability power of the Momentum Factor (UMD, as defined in Appendix \ref{app2}, on
    the European stock pool, 1985-2020). For every day and every stock in the pool,
    we put on the X-axis the value of the descriptor properly
    normalized, and on the Y-axis the future residual return of
    the corresponding stock (i.e. the total stock return minus its beta
    component on the largest principal component of the covariance matrix). All these points are then averaged inside bins (the standard errors for these averages are also represented). The plain line is a linear regression through the points with a positive predictor. The dashed line shows the 40\% threshold that short predictors have
    to beat in order for shorts to be accretive, according to the model.}
  \label{predplot}
\end{figure}

\begin{figure} 
  \centering
\includegraphics[width=.8\textwidth]{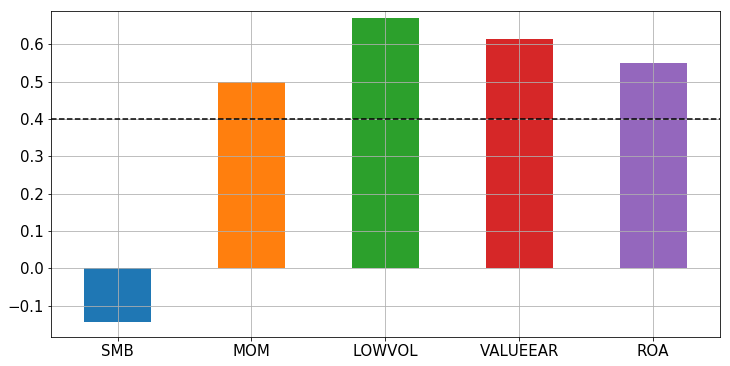}
  \caption{Global (USA-Canada, Europe, Japan, Australia, 2000-2020) view of the predictability plot of Fig.~\ref{predplot}. \red{For the factors SMB, MOM, LOWVOL, VALUEEAR and ROA (sample and methodology are described in Section \ref{toymodel}), on a world-wide pool of stocks,}  we show the ratio of the slope of the points corresponding to negative values of the predictor to the slope of the points corresponding  to positive values of the predictor. \red{The dashed line materializes the 40\% threshold of our toy model beyond which $\text{SR}(\pi_{\text{LS}})>\text{SR}(\pi_{\text{LH}})$. It appears that for all factors but SMB, the short leg under-performs the market enough to justify the long-short implementation.}}
  \label{pred_plot_lin_reg_ratios}
\end{figure}

Our simple model suggests a quantitative threshold on the predictability of
the short vs. the long positions beyond which a Long-Short implementation is beneficial. 
This threshold can be used on real data as we show
in Figs.~\ref{predplot},~\ref{pred_plot_lin_reg_ratios}. 
\red{The sample of the global market we use is a a monthly re-balanced world-wide pool made of the 1200 (resp. 1000, 900, 200) most liquid stocks (liquidity being quantified as the 180-days Average Daily Traded Volume) of North-America (resp. Europe, Japan, Australia), from 2000 to 2020, using the definition of the Equity Factors we give in Appendix \ref{app2}.}

The data reveals that for most of the equity
factors that we will consider in this study (see next section), the
short side of the predictor is sufficiently strong to make
short positions profitable, at least in principle. In other words, taking short equity positions is \emph{a priori} a better choice than simply shorting the future contract. 
If the residual returns $\varepsilon_i$ have the same volatility
($\kappa=1$), then as soon as $\alpha_2>\sqrt{2}-1$ (an
under-performance of Asset 2 with respect to Market at least
40\% of the out-performance of Asset 1), we find 
$\text{SR}(\pi_{\text{LS}})>\text{SR}(\pi_{\text{LH}}).$ This threshold is
materialized by dashed lines in Figures ~\ref{fig_sr_ratio},~\ref{predplot},~\ref{pred_plot_lin_reg_ratios}.

Finally, note that since $\sqrt{1+\kappa}-1$ is an increasing function of
  $\kappa$, the higher the volatility of the residual returns of
  Asset 2, the stronger the under-performance of Asset 2 \red{needs to be} for $\text{SR}(\pi_{\text{LS}})>\text{SR}(\pi_{\text{LH}})$ to hold.

\section{Revisiting The Paper \red{\emph{When Equity Factors Drop Their Shorts}}}\label{parsdestruens}

In this section we revisit the results reported in ref. \cite{Robeco2}. The study is based on the classical Fama-French factors, as available in the Kenneth French data
library~\footnote{https://mba.tuck.dartmouth.edu/pages/faculty/ken.french/data\_library.html}. We
consider here the following factors: HML (aka Value), WML (aka Momentum), RMW (aka Profitability), CMA (aka Investment), and also VOL, i.e. a low volatility factor available online~\footnote{https://www.robeco.com/en/themes/datasets/}. These
factors are built using monthly returns of so-called Fama-French ``2x3'' portfolios on the US market (see the corresponding websites for more details). This construction allows one to clearly distinguish a long and a short leg for each factor. 

\begin{figure} 
  \centering
\includegraphics[width=.45\textwidth]{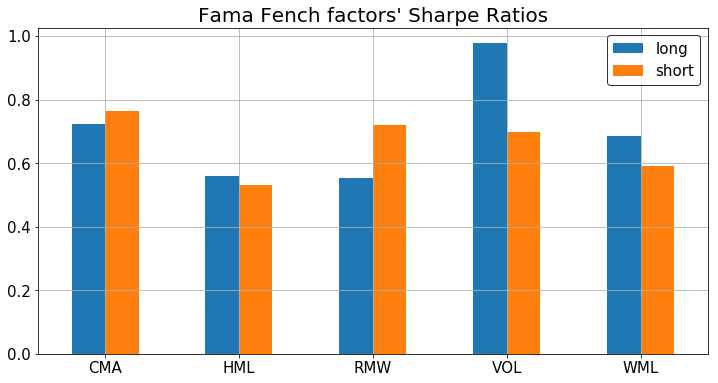}\qquad\includegraphics[width=.45\textwidth]{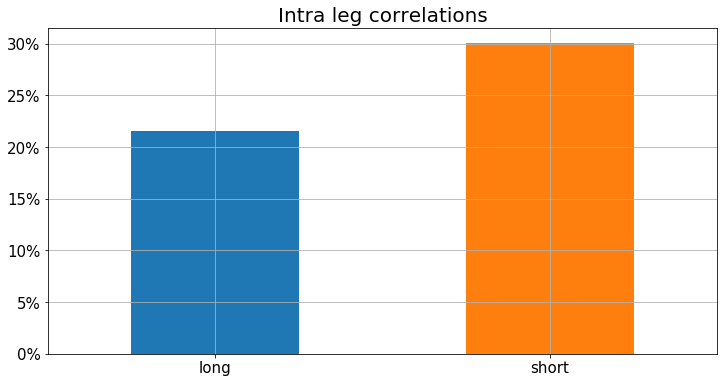}
  \caption{\textbf{Fama-French portfolios.} \textit{Left:} Sharpe Ratios of hedged long and short legs. \textit{Right:} mean correlation, across factors, of long (resp. short) legs with other long (resp. short) legs.  These measures result in  a portfolio with  maximum Sharpe (based on the 10 synthetic assets defined by the hedged long and short legs of these 5 factors) weighting long (resp. short) legs at about $70\%$ (resp. $30\%$).}
  \label{FF_long_vs_shorts}
\end{figure} 

In order to run a fair comparison between long and short legs, 
one beta-hedges each factor using a ``market index'' \red{(namely the Fama-French market)} built using the same 2x3 building blocks. Note that by construction this index contains $50 \%$ of small cap. stocks and $50 \%$ of large cap. stocks. Each factor is rescaled to get a beta of one with respect to the index. The index contribution is then removed to get a beta-neutral leg. 

    \begin{figure} 
  \centering
\includegraphics[width=.6\textwidth]{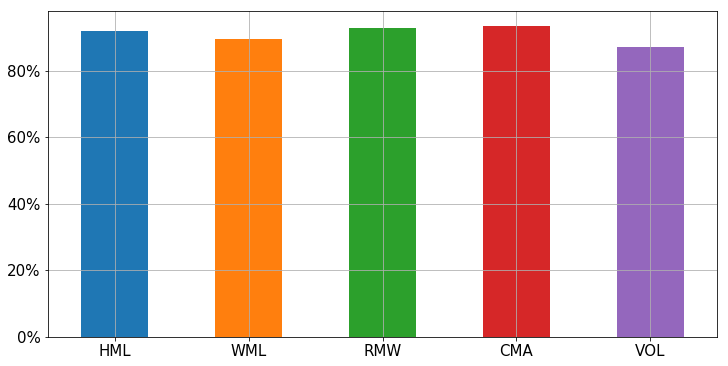}
  \caption{\textbf{Fama-French portfolios.} \red{This chart shows that when \emph{(longs $-$ market)} outperforms \emph{(market $-$ shorts)}, this is in large part explained by an SMB exposition of the difference \emph{(longs $-$ market)} $-$ \emph{(market $-$ shorts)}.} We show here the correlation, for each factor, between SMB and the difference $\Delta=${(longs $-$ $\beta_{\text{longs}}$SPmini)} $-$ {($\beta_{\text{shorts}}$SPmini $-$ shorts)}.
This correlation can be explained as follows: $\Delta$ roughly rewrites as {(longs $+$ shorts) $-$ 2 SPmini}. Given the Fama-French portfolios used in \cite{Robeco2}  are 50\% small caps and 50\% large caps, $\Delta$ is correlated to the difference between and equally-weighted index with a market-cap weighted index, i.e. to SMB.} 
  \label{long_vs_shorts_corr_with_smb}
\end{figure} 

The Sharpe ratio of each beta-neutral leg is shown in Fig.~\ref{FF_long_vs_shorts}-Left. Overall, there is no striking difference of Sharpe ratio between the long and short legs, except perhaps for VOL. However,  
Fig.~\ref{FF_long_vs_shorts}-right reveals that the returns of the different short legs are significantly more correlated than that of the long legs, as pointed out in \cite{Robeco2}. Hence, an optimal allocation will be more tilted towards the long portfolios although, at variance with \cite{Robeco2}, we find that the short leg should be allocated $30 \%$ of the weight, and not zero weight. The discrepancy between our conclusions may be related to subtle implementation differences. \red{We have not been able to identify precisely which implementation detail matter most, but we suspect the detailed definition of the optimization problem is relevant. For example, there are different ways to estimate the correlation structure of the factors, and various constraints that one may want to impose to the portfolio.}
At the very least, this means that the ``no-short'' recommendation is not robust against such minor changes. 

More importantly, the very definition of the market index -- which serves as a hedge -- is found to matter quite dramatically. Using for example the SPmini index (which is easily implementable as a low-cost hedge), we now find that the Sharpe ratio of the long legs is significantly better that that of the short legs, as reported in \cite{Robeco2}. However, this is because the difference between the two legs is now mechanically exposed to the SMB factor: see Fig.~\ref{long_vs_shorts_corr_with_smb}. 

We now turn to the analysis of a realistic implementation of factor trading, including the important issue of costs, in both its Long-Short (LS) and Hedged Long-Only (LH) incarnations. Our main conclusion will be that factor trading through Long-Short portfolios actually over-performs Hedged Long-Only portfolios.

\section{A Realistic Implementation Framework}\label{parsconstruens}
\subsection{Description}
We define a set of predictors based on ranked metrics (\textit{Low Volatility, Momentum, Returns Over Assets, Small Minus Big,  Value Earnings\footnote{\red{The version of Value based on the Book/Price ratio is excluded because of its long time bad performance.}}}, see Appendix \ref{app2}) which, we believe, represent well the equity factor space, on a pool of stocks distributed over the main exchanges (USA-Canada, Europe, Asia, Australia), proportionally to the available liquidity. We prefer ranked implementations over 2x3 implementations since, as shown for example in Fig. \ref{predplot}, there is predictability even within the ``central'' quantiles, not only the extremes one. Our study spans the period 2000-2020.   

We then slow down the corresponding signals (via an Exponential Moving Average of 150 days), so that the turnover of the portfolios reach reasonable values, compatible with typical trading costs. More precisely, the turnover  is $\approx 0.5\%$ of AUMs per day for the LH portfolio and $\approx 2\%$ of the Gross Market Value for the LS portfolio.

The LH portfolio is implemented via a long-only constrained optimization problem where the portfolio's overlap with the factor predictor (or of the aggregated factors) is maximized, at controlled turnover cost. More precisely, the portfolio is updated daily as the solution  of the  optimization problem 
$$\text{portfolio}_t=\operatorname{Argmax}\left(\langle\text{portfolio}_t \cdot \text{factors predictor}_t\rangle - \text{trading costs}\right),$$ 
under global AUM constraints plus individual risk constraints (i.e. a maximum position \red{of $3\%$ of the AUMs} on every single stock). 

The trading costs are the sum, for each trade, of a linear term accounting for bid-ask spread and broker costs, plus a term accounting for market impact that depends on the corresponding stock's liquidity. These trading costs are computed using our best in-house estimates of all these separate contributions, in particular of the square-root impact law documented in, e.g. \cite{Bucci_Ancerno} and refs. therein.  

   \begin{figure} 
  \centering
\includegraphics[width=.45\textwidth]{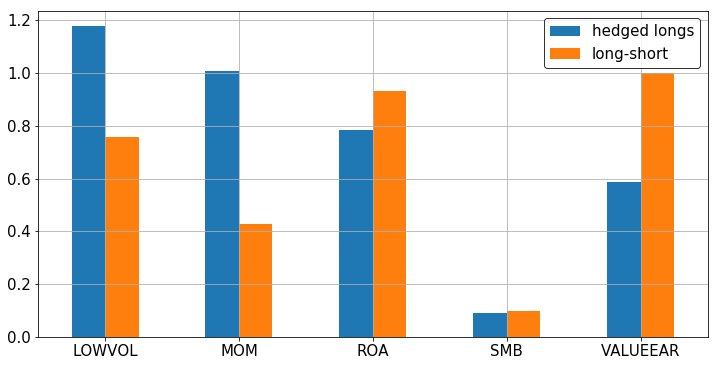}\qquad\includegraphics[width=.45\textwidth]{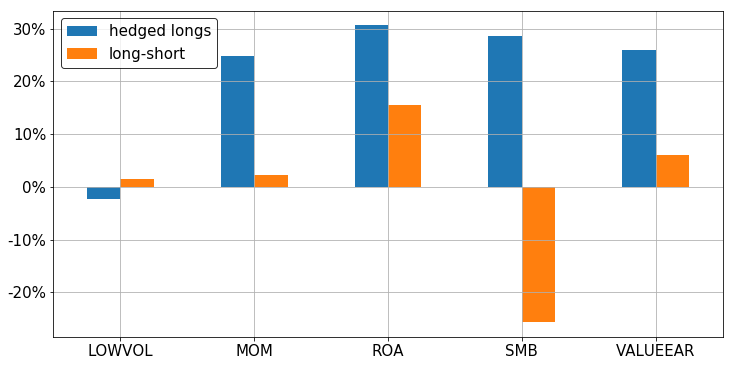}
  \caption{\textit{Left:} Sharpe Ratios of the LH and LS implementations of the different factors. \textit{Right:} Mean correlation of each factor with the other factors in the LH and LS implementations.} 
  \label{LSLH_factors_sharpes}
\end{figure} 

We then look at the tracking error of the long-only portfolio with respect to easily tradeable market indexes (respectively: S\&P 500, DJ EURO STOXX 50, TOPIX, S\&P ASX 200 and S\&P/TSX 60 Total Return indices). This tracking error corresponds to our LH implementation. The LS portfolio is constructed using the same signals and the same cost control, plus a volatility target equal to that of LH, \red{implemented using a ``cleaned'' version of the empirical stock returns correlation matrix, see  \cite{BouchaudPotters}}. This construction requires a certain leverage ratio that leads to some extra financing costs, together with some idiosyncratic shorting costs for \textit{hard-to-borrow} stocks, which we carefully account for\footnote{We use our in-house data base for hard-to-borrow fees, which corresponds to the fees actually paid by the CFM equity market neutral programs over the years}. 

\begin{figure} 
  \centering
\includegraphics[width=.6\textwidth]{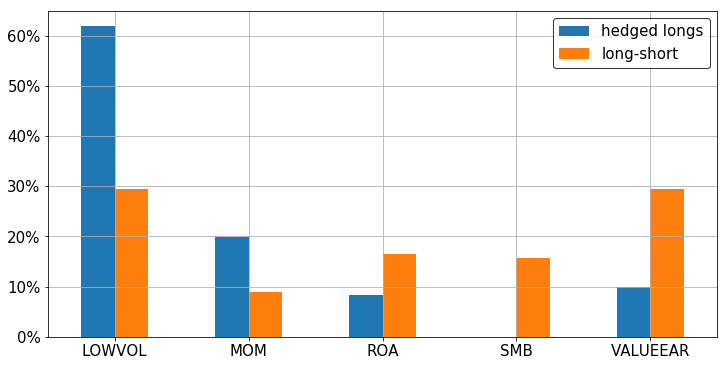}
  \caption{Factors' weights in maximum Sharpe LH and LS portfolios.} 
  \label{max_sharpe_ptf_LSLH}
\end{figure} 

\begin{figure} 
  \centering
\includegraphics[width=.48\textwidth]{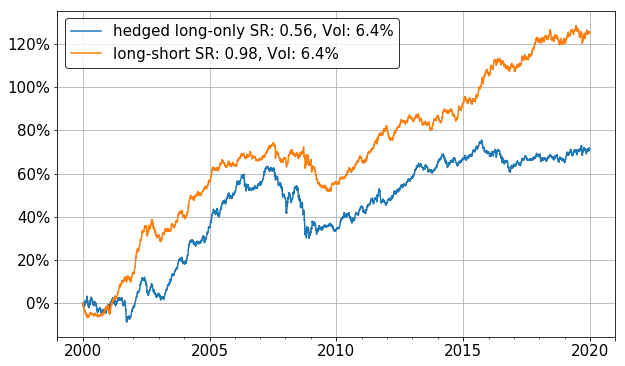}\quad \includegraphics[width=.48\textwidth]{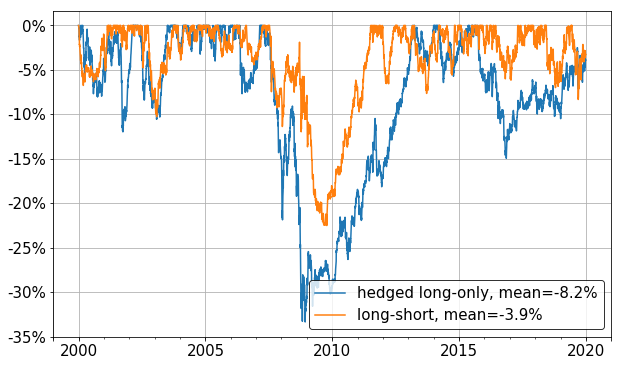}
  \caption{Left: P\&L (including all costs) of the LH and LS implementations, both with a realised volatility of $6.4\%$ annual. Both portfolios take as an input signal the equi-weighted sum of all factors predictors: \textit{Low Volatility, Momentum, Returns Over Assets, Small Minus Big, Value Earnings}. The Sharpe ratio of LH/LS are, respectively 0.56 and 0.98. Right: Drawdown time series (${\text{AUM}-\text{Peak Value}}/{\text{AUM}}$) for both implementations. The mean depth for LH amounts to -8.2\%, but only -3.9\% for LS.} 
  \label{5_factors_signal_blending_LS_vs_HLO_pnl}
\end{figure}

The results of these experiments, using either single factors or the aggregate of different factors, are summarized below and illustrated in Figs.~\ref{LSLH_factors_sharpes}, ~\ref{max_sharpe_ptf_LSLH} and  
\ref{5_factors_signal_blending_LS_vs_HLO_pnl}. We fix the size of the AUMs to 1 Billion USD.

\begin{itemize}
\item First, we see that the LH implementation of individual factors leads to Sharpe Ratios that are comparable, on average, to those of the LS implementation (see Fig.~\ref{LSLH_factors_sharpes}, \textit{left}). But we now see that the LH implementations of the different factors are way more correlated than their LS counterparts (see Fig.~\ref{LSLH_factors_sharpes}, \textit{right}). 
 
\item Taking the factors' Sharpe ratios and their correlations into account, we compute the weight of each factor in the  maximum Sharpe portfolios for both cases (LH and LS). We see, as expected, that the LS implementation is much more diversified between the different factors (see Fig.~\ref{max_sharpe_ptf_LSLH}). Note that better diversification allows one to expect, as a general rule of thumb, a more robust out-of-sample performance. 

\item Finally, the P\&Ls of the LH and LS implementations of aggregated factors (with equal weights) are presented in Fig.~\ref{5_factors_signal_blending_LS_vs_HLO_pnl} (left). The Sharpe ratio of the LS implementation is found to be $\approx 1$, significantly higher than the one of the LH implementation ($\approx 0.56$). Let us emphasize that {\it all} costs (trading costs, leverage financing costs of LS, and borrowing costs of LS) are included in these P\&L. The breakdown of these costs is detailed in Table~\ref{tab:LS_vs_LO}. Note that although total costs are $\approx 60 \%$ higher for the LS implementation, it still significantly over-performs the LH implementation. 
\end{itemize}

\subsection{Effect Of The AUM Level}
\subsubsection{Market Impact}
  Note  that the conclusion above may not hold for very large AUMs, given the higher turnover of the LS implementation and the super-linearity of impact costs as a function of trading volume. Indeed, according to e.g. \cite{YCostPaper}, when ignoring the second order effect of the slow decay of the impact described in \cite{Bucci_Ancerno}, the price $P_{\text{effective}}$ at which one can buy (resp. sell)  shares of a stock $s$ with spot price $P$ can be modeled as follows: \begin{equation}\label{trade_cost}P_{\text{effective}}=P+\varepsilon \left(\frac{\text{average bid-ask spread}}2+\text{broker fees}+\text{impact cost}\right)\end{equation}  where $\varepsilon=1$ (resp. $-1$) and where the impact cost term can be modeled as 
  \begin{equation}\label{y_cost}\text{impact cost}=Y \times P \times \operatorname{Volatility}(s)\sqrt{\frac{Q}{\operatorname{ADV}(s)}},\end{equation} for $Y$ a constant of order $1$, $Q$ the traded volume and $\operatorname{ADV}(s)$ the average daily volume of $s$ traded on the market.
  
  In Equation \eqref{trade_cost}, the average bid-ask spread term and the broker fees term have effects that are linear in the traded volume, hence in the AUMs. The super-linearity comes from the impact term. Let us compute the associated cost for a strategy that trades daily a fraction $q$ of its AUMs, distributed on $m$ stocks\footnote{We can suppose for example that $m\approx 10^3$ if the trading universe has $5000$ stocks, each traded weekly. Indeed,  at large AUMs, \eqref{y_cost} incites to trade rather frequently than once in a row to update positions.}). 
  Introducing $\mathcal{M}$ as the median value of $\frac{\operatorname{Volatility}(s)}{\sqrt{\operatorname{ADV}(s)}}$ on the trading universe, we get $$\text{daily total impact cost}\approx mY\mathcal{M}\left(\frac{q\text{AUMs}}m\right)^{3/2}=Y\mathcal{M}\frac{\left(q\text{AUMs}\right)^{3/2}}{\sqrt{m}}.$$
  
  We see that in the case of the LS portfolio, for which $q$ is larger than for the LH portfolio, the super-linear effect of the impact is stronger, and ends up reversing its initial advantage.
  
  \subsubsection{Hard To Borrow Stocks}
  
  Stocks cannot be shorted without limit, and therefore hedge funds can only utilize a limited amount of short positions for each given stock. As the AUMs of a fund manager taking short positions increase, this constraint is getting more stringent and impair the optimal implementation of factor investing. In contrast, shorting important amounts of the market index using futures is relatively easy.  These effects therefore also reduce the advantage of the LS portfolio over the LH one as the AUMs increase.
  
  \begin{table}
  \begin{center} \begin{tabular}{| c || c | c | c | c | c |  c |}
 \hline     
&  Sharpe & mean drawdown & returns+div. & trad. cost & financ. cost & short borrow. cost \\  \hline\hline 
  LH & 0.56 & -8.2\% & 8.4\% & -2.8\%  & -2.0\% & \text{NA}  \\ \hline   
  LS & 0.98 & -3.9\% & 14.2\%  & -4.8\%  & -2.6\% & -0.6\%  \\
 \hline
  \end{tabular}
  \end{center}
  \caption{LH vs LS main P\&L statistics, with returns and costs in percent per year of the 1 Billion USD AUMs.}
    \label{tab:LS_vs_LO}   
\end{table}
  
\section{Conclusions}\label{conclusions}

The conclusions of our study are quite clear: first, when discussing the relative merits of Hedged Long-Only and Long-Short portfolios, details matter. One should carefully discuss what ``market'' is used to hedge the Long-Only positions, since different definitions can lead to uncontrolled exposures to some factors, such as Small Minus Big or Low Volatility. All implementation costs should be estimated and integrated in the final P\&L horse-race. Once all this is done, and provided our analysis is error-free, we unambiguously find that Long-Short implementations best Hedged Long-Only ones, at least when the AUMs are not too large. This conclusion is bolstered by the analysis of a very simple toy model, which provides a threshold on the strength of short predictors. Empirical predictability of the shorts indeed seem to lie beyond that threshold by a substantial margin. 

\vskip 1cm

\noindent\emph{Acknowledgments.} We thank  M. Cristelli, J.C. Domenge, S. Gualdi, T. Madaule, P. Seager and S. Vial for interesting discussions and suggestions.

\appendix
\section{More On The Toy Model}\label{app1}

In this appendix, we show how we get equation~\eqref{eq_ratio_sr} in
Section \ref{toymodel}. We have
\begin{equation}
  \text{PNL}(\pi_{\text{LS}})=(1+\alpha_2)F+\varepsilon_1-\varepsilon_2 \ ,
\end{equation}
so that
\begin{equation}
  \text{SR}(\pi_{\text{LS}}) = \frac{(1+\alpha_2)\mathbb{E}(F)}
       {\sqrt{(1+\alpha_2)^2\text{Var}(F)+\text{Var}(\varepsilon_1)+\text{Var}(\varepsilon_2)}}
\end{equation}
Similarly,
\begin{equation}
  \text{SR}(\pi_{\text{LS}})=\frac{ \mathbb{E}(F)}{\sqrt{
      \text{Var}(F)+\text{Var}(\varepsilon_1)}}
\end{equation}
Eq.~\eqref{eq_ratio_sr} follows immediately and we deduce that 
\begin{eqnarray}
  \text{SR}(\pi_{\text{LS}})>\text{SR}(\pi_{\text{LH}})
  &\iff&(1+\alpha_2)^2>1+\kappa  \\ &&\\
  &\iff& \alpha_2>\sqrt{1+\kappa}-1,
\end{eqnarray}
where we used the hypothesis that $1+\alpha_2>0$.

\section{Brief Description Of The Equity Factors}\label{app2}
\begin{itemize}
\item Momentum: 11 month mean of returns lagged by 1 month ranked
\item Value Earnings: earnings/price ranked 
\item Low Volatility: 250 day volatility anti-ranked
\item Size: market cap (lagged by  20 days, averaged over 40 days) ranked
\item ROA: net income/total assets ranked
\end{itemize}


\begin{thebibliography}{99}

\bibitem{Amihud2002} Y. Amihud  (2002), \emph{Illiquidity and stock
  returns: cross-section and time-series effects,}  Journal of Financial Markets 5 (1), 31-56.

\bibitem{Baker2011} M. Baker, Malcolm, B. Bradley and J.
  Wurgler (2011), \emph{Benchmarks as Limits to Arbitrage: Understanding
  the Low-Volatility Anomaly,} Financial Analyst Journal, Vol. 67,
  No. 1, pp. 40–54.

\bibitem{Bambaci} J. Bambaci, J. Bender, R. Briand, A. Gupta, B. Hammond and 
M. Subramanian (2013), \emph{Harvesting Risk Premia for Large Scale Portfolios,}
MSCI report
\url{https://www.regjeringen.no/contentassets/f453b48778c342d9a7ed8d6810d6cea4/harvesting_risk.pdf}

\bibitem{Robeco2}  D. Blitz, G. Baltussen and P. van Vliet
  (2020), \emph{When Equity Factors Drop Their Shorts},  Financial Analysts Journal. 

  
  \bibitem{Bouchaud2016}
  J.-P. Bouchaud, P. Krueger, A. Landier and D. Thesmar (2018),
\emph{Sticky Expectations and the Profitability Anomaly}, Journal of Finance

\bibitem{BouchaudPotters} J.-P. Bouchaud, M Potters (2003), \emph{Theory of financial risk and derivative pricing: from statistical physics to risk management}
Cambridge university press

  
\bibitem{Bucci_Ancerno} F. Bucci, M. Benzaquen, F. Lillo and J.-P. Bouchaud (2019), \emph{Slow decay of impact in equity markets: insights from the ANcerno database}.  \url{https://arxiv.org/abs/1901.05332}
  
\bibitem{Carhart} M.M. Carhart (1997), \emph{On Persistence in Mutual Fund Performance}, The Journal of Finance {52}, 57 
  
\bibitem{Chan1991} K.C. Chan and N.F. Chen  (1991),
  \emph{Structural and return characteristics of small and large
    firms,}  Journal of Finance 46: 1467–84.
    
\bibitem{Dasgupta2011} A. Dasgupta, A. Prat and M.
  Verardo (2011), \emph{The price impact of institutional herding,}
  Review of Financial Studies, 24 (3): 892-925.
  
      \bibitem{DeBondt1987} W.F.M. De Bondt    and R.H. Thaler
  (1987), \emph{Further Evidence on Investor Overreaction and Stock
  Market Seasonality,}  Journal of Finance. 42:3, pp. 557-81.

\bibitem{FamaFrench93} E.F. Fama and K.R. French (1993), \emph{Common risk factors in the returns on stocks and bonds}, Journal of Financial Economics, {33}, 3 


\bibitem{HarveyLiu2019} C.R. Harvey and Y. Liu (2019), \emph{A Census of the
  Factor Zoo,}
  \url{https://papers.ssrn.com/sol3/papers.cfm?abstract_id=3341728} 

\bibitem{HarveyLiu2013} C.R. Harvey, Y. Liu and C. Zhu (2013), \emph{...and
  the Cross-Section of Expected Returns,}
  \url{https://papers.ssrn.com/sol3/papers.cfm?abstract_id=2249314}
  
  
  \bibitem{Robeco1} J. Huij, S. Lansdorp, D. Blitz and P. van Vliet (2014),
\emph{Factor Investing: Long-Only versus Long-Short}, 
\url{https://papers.ssrn.com/sol3/papers.cfm?abstract_id=2417221}
  
  \bibitem{Ilmanen2012} A. Ilmanen and J. Kizer  (2012), \emph{The Death
  of Diversification Has Been Greatly Exaggerated},
  \url{https://papers.ssrn.com/sol3/papers.cfm?abstract_id=2998754}

\bibitem{Ross1976} S. Ross (1976), \emph{The arbitrage theory of
  capital asset pricing}, Journal of Economic Theory 13(3), 341–360
  
\bibitem{Sharpe1964} W. Sharpe (1964), \emph{Capital Asset Prices:
    A Theory of Market Equilibrium under Conditions of Risk,}  Journal of Finance 19, 425-442.
  
  \bibitem{Sloan1996} R. Sloan (1996), \emph{Do stock prices fully
  reflect information in accruals and cash flows about future
  earnings?}  The Accounting Review 71: 289-315.

\bibitem{Stambaugh2011} R.F. Stambaugh, J. Yu and Y. Yuan
  (2011), \emph{The Short of It: Investor Sentiment and Anomalies},
\url{https://papers.ssrn.com/sol3/papers.cfm?abstract_id=1567616}  


\bibitem{YCostPaper} B. T\'oth, Y. Lemperiere, C. Deremble, J. De Lataillade, J.Kockelkoren and J.-P. Bouchaud (2011), \emph{Anomalous price impact and  the  critical  nature  of  liquidity  in  financial  markets}. Physical Review X, vol. 1, no. 2, p. 021006.
  
\bibitem{Vassalou2004} M. Vassalou and Y. Xing (2004), \emph{Default
  Risk in Equity Returns,}  Journal of Finance 59, 831-868.
  
\bibitem{Zhang2006} X.F. Zhang (2006), \emph{Information uncertainty
  and stock returns,} Journal of Finance 61(1), 15-136.




  
\end{thebibliography}
\end{document}